\documentclass[fleqn,twoside]{article}
\usepackage{espcrc2}

\usepackage{graphicx}
\usepackage[figuresright]{rotating}

\newcommand{\be}{\begin{eqnarray}}
\newcommand{\ee}{\end{eqnarray}}
\newcommand{\bi}{\bibitem}

\newcommand{\rar}{\rightarrow}
\newcommand{\lrar}{\leftrightarrow}
\newcommand{\nue}{\nu_e}         
\newcommand{\num}{\nu_\mu}
\newcommand{\nut}{\nu_\tau}

\newcommand{\AmS}{{\protect\the\textfont2
  A\kern-.1667em\lower.5ex\hbox{M}\kern-.125emS}}

\title{Big Bang Nucleosynthesis
}

\author{A.D. Dolgov \address{
INFN, sezione di Ferrara, Via Paradiso, 12 - 44100 Ferrara,
Italy}
\address{
ITEP, B. Cheremushkinskaya 25, Moscow, 117259, Russia
}}

\begin{document}

\begin{abstract}
A review of Big Bang Nucleosynthesis (BBN) is presented. Observations of 
deuterium and helium-4 are discussed. Some BBN restrictions on non-standard
physics, especially on neutrino properties and time-variation of fundamental 
constants are given.
\vspace{1pc}
\end{abstract}

\maketitle

\section{INTRODUCTION \label{s-intr}}

Big Bang Nucleosynthesis (BBN) is known to be one of three solid pillars on
which the Standard Cosmological Model (SCM) stands. 
The other two include General Relativity (GR) 
and Cosmic Microwave Background Radiation (CMBR). 
An agreement of BBN calculations of light element abundances
with observations presents the strongest proof in favor of the statement
that 12-14 billion years ago
the universe was indeed hot with the temperatures in MeV range and that
the entropy per baryon is huge, about $10^{9}$.

According to the theory, light elements $^2H$, $^3He$, $^4He$, and $^7Li$
have been created in the early universe during first few hundred seconds of
her existence. The abundances of these elements span 9 orders in magnitude
and are in excellent, good, or reasonable agreement with the observational 
data, depending upon the moment when the comparison of theory with the data
is taken and upon the personal point of view of a researcher.

The theory of BBN is robust, well defined, and quite precise. The largest 
uncertainty is introduced by the values of the cross-sections of nuclear
reactions. Theoretical accuracy is better than 0.1\% for $^4He$, better
than 10\% for $^2H$ and is about 20-30\% for 
$^7Li$~\cite{fiorentini98,burles99}. In all the cases theoretical
uncertainty is 
much smaller than observational precision. Observations of light elements
encounter two serious problems: systematic errors and evolutionary effects.
We will discuss them below.

In the next section physics of BBN and essential parameters and inputs
are described. In section 3 observational data are analyzed (looking from
outside by a non-expert). In section 4 modifications of the standard
scenario are discussed. Conclusion is presented in the
last section 5.

\section{PHYSICS OF BBN \label{s-physbbn}}

Physical processes which were essential for cosmological creation of light 
elements took place
at the temperatures, $T$, in the range of a few MeV down 
to 60-70 keV. The corresponding time, $t$, interval was from 
a few$\times 0.1$ sec up to 
$10^3$ sec. At this time the universe was dominated by relativistic matter 
and the rate of cosmological cooling was determined by the expression:
\be
\left({t\over {\rm sec}}\right )\left({T\over {\rm MeV}}\right)^2 
= 0.74 \left( {10.75 \over g_*}\right)^2
\label{tT2}
\ee
where $g_*$ is the effective number of particle species in the cosmic plasma. 
In the standard model the factor
$g_*$ includes contributions from photons equal
2, from $e^\pm$-pairs equal 7/2, and the contribution from three neutrino
flavors equal $3\cdot 7/4$. Any additional, non-standard form of energy is 
parametrized in terms of effective number of neutrino species:
\be
g_* = 10.75 + {7\over 4}\,(N_\nu -3)
\label{gstar}
\ee
Since the cosmological cooling rate depends upon $g_*$ it is clear that
BBN is sensitive to {\it any form of matter/energy} present in the primeval
plasma in the temperature interval ($\sim$ MeV) - 0.06 keV. This effect was
noticed in refs.~\cite{hota64}; detailed calculations of the effect
were pioneered in papers~\cite{shvartsman69}. 

The first step in creation of light elements is ``preparation'' of neutrons.
Their number density is determined by the reactions:
\be
n + e^+ &\lrar& p + \bar \nu_e, \nonumber \\
n +\nu_e &\lrar& p + e^-
\label{np-react}
\ee
Since the reaction rate is proportional to $T^5$ and the expansion rate
is $H\sim T^2$ thermal equilibrium is maintained at high temperatures
and $n/p$-ratio follows the equilibrium curve:
\be
n/p = \exp [-(\delta m + \mu_e )/T]
\label{n/p}
\ee
where $\Delta m = 1.293$ MeV is the neutron-proton mass difference and
$\mu_e$ is the chemical potential of electronic neutrinos. In the standard 
model the latter is assumed to be vanishingly small. 

At $T= 0.6-0.7$ MeV expansion became faster than reactions (\ref{np-react})
and the $n/p$-ratio would tend to a constant, if not the neutron decay. 
Because
of the decay the ratio slowly decreases as $\exp (-t/\tau_n)$ with
$\tau_n = 889$ sec. This behavior lasts approximately till $T_{NS}=65$ keV 
when light element formation abruptly begins and all free neutrons quickly
disappear from the plasma forming deuterium, helium, and lithium.
The value of $T_{NS}$ is determined by the binding
energies of light nuclei, in particular deuterium, and by the 
baryon-to-photon ratio, $\eta = n_B/n_\gamma$. 

Practically all neutrons that remained in the plasma to the moment when the
temperature dropped down to $T_{NS}$ ended their lives in
$^4He$ because the latter has the largest binding energy. The mass fraction 
of the primordial $^4He$ should be about 25\%. A little deuterium and $^3He$
survived because they were not able to find a nucleon, $N=p,n$, for further 
transformation into $^4He$ through the chain of reactions: 
\be
N + {^2H} \rar (^3H,\,\, ^3He) \nonumber \\
(^3H,\,\, ^3He) + N \rar ^4He
\label{nucl-react}
\ee
As a result the abundance of primordial deuterium would quickly decrease with 
rising baryon number density or, better to say, parameter $\eta$, and its
output with respect to hydrogen would be (a few)$\times 10^{-5}$ by number. 
A strong sensitivity of deuterium abundance to the baryon-to-photon ratio
$\eta$ makes this element a very convenient tool to measure the baryonic
charge of the universe. That's why primordial deuterium is called 
``baryometer'', the term suggested by David Schramm.

An absence of a (quasi)stable nuclei with $A=5$ inhibits production of
$^7 Li$, because the latter should be produced by fusion of two lighter
nuclei and not in a sequence of reactions of free protons or neutrons with
nuclei. Correspondingly the fraction of $^7 Li$ is very small,
(a few)$\times 10^{-10}$.

\begin{figure}[htb]
\vspace{9pt}
\includegraphics[scale=0.35]{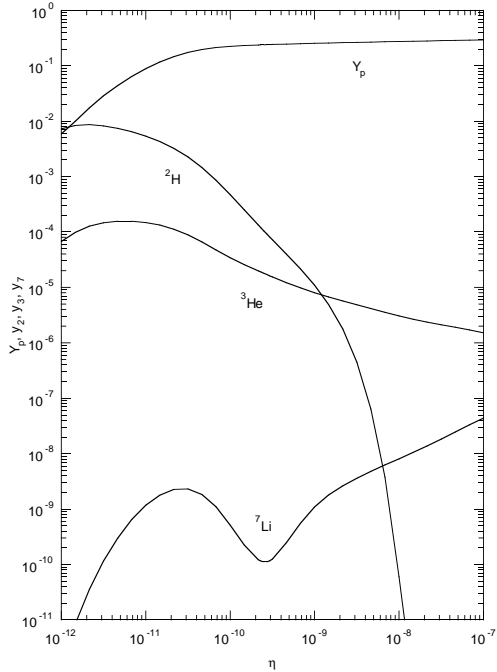}
\caption{The predicted abundances of light elements as a function of 
$\eta_{10}$.
}
\label{elmnt}
\end{figure}

The abundances of light elements as functions of the baryon number density,
$\eta_{10} = 10^{10} n_B /n_\gamma$ are presented in fig.~\ref{elmnt},
taken from the review~\cite{olive00ss}. 

As we have seen, the abundances of light elements essentially depend upon
the following parameters:
\begin{enumerate}
\item{}
Baryon-to-photon ratio, $\eta_{10} = 10^{10} n_B /n_\gamma$. Only two years 
ago this parameter was determined from BBN itself but now there is an
independent way to measure it through the spectrum of angular fluctuations
of CMBR. According to the BOOMERANG and DASI measurements:
\be
\Omega_B h^2 = \left\{ \begin{array}{ll}
0.021 \pm 0.004 & \mbox{\cite{boom}},  \\
0.022^{+0.004}_{-0.003} & \mbox{\cite{dasi} },
 \end{array}\right.
\label{omegab}
\ee
where $\Omega_B = \rho_B /\rho_{crit}$ is the fraction of baryon mass density
in terms of cosmological critical energy density and 
$h = H/100$km/sec/Mpc is the dimensionless Hubble parameter. 
$\Omega_B h^2 = 0.022$ corresponds to $\eta_{10} = 5.7$. A higher value
is given by MAXIMA-I measurements~\cite{maxima}, 
$\Omega_B h^2 = 0.0325\pm 0.007$ which disagrees with the other groups by 
two standard deviations.
\item{}
Rate of the reactions (\ref{np-react}). It is expressed through neutron
life-time which is pretty well known now, $\tau_n = 886.7\pm 1.9$~\cite{pdg}.
\item{}
Total cosmological energy density during BBN. This quantity is usually 
parametrized as the number of additional neutrino families,
$\Delta N = N_\nu -3$. This is a precise parametrization if the additional
energy is in the form of relativistic particles with the equation of state
$p=\rho/3$. 
However it is not so for matter with a different equation of state,
e.g. nonrelativistic matter or vacuum-like energy. In this case the variation
of different abundances would be different from that induced by extra
neutrinos. 
\item{}
Neutrino degeneracy. It is assumed usually that the charge asymmetry of
different neutrino flavors is vanishingly small. Thus their chemical
potentials are zero or negligible. Strictly speaking their values are 
unknown and the best way to determined them or obtain an upper bound on
their magnitude is BBN. Chemical potentials of $\nu_\mu$ and $\nu_\tau$
can be described by $\Delta N$, since their role is only to increase the
energy density of relativistic matter at BBN. Element abundances are much
more sensitive to chemical potential of $\nu_e$ because the latter not only
changes the canonical energy density but also directly shifts the 
$n/p$-ratio~(\ref{n/p}).
\end{enumerate}

\section{OBSERVATIONAL DATA \label{s-obs} }
\subsection{Observations of Deuterium \label{ss-deuterium}}

The present day abundance of deuterium can be noticeably different from the
primordial value because deuterium can be burnt in stars producing $^3He$.
Exact evolutionary effects are uncertain. One can rigorously say only that
the observed today deuterium presents a lower bound to the primordial one.
For more detail see e.g. review~\cite{olive00ss}.
Fortunately in recent years observation of deuterium in large $Z$ low
metallicity ($<1/50$ of the solar) clouds of cold neutral gas (HI) became
possible~\cite{carswell94,songalia94,tytler94,webb97,burles98}.
Such clouds are presumably not contaminated by stellar
processes and the observed deuterium may be the primordial one;
see however ref.~\cite{prantzos01} where this statement is questioned.

Since the deuterium isotope shift corresponds to velocity of only 
(-82) km/sec the clouds with a simple velocity structure are 
necessary for reliable data. 
In addition to the uncertainty induced by the unknown velocity structure,
ionization corrections and a possible ``interloper'' further increase the
systematic errors. A discussion of these effects and a list of references 
can be found e.g. in~\cite{tytler00,5}. 

Because of these problems, there are only a few measurements of 
deuterium at high red-shifts available now. They are summarized in
Table~\ref{t-h2}.

\begin{table*}[htb]
\caption{
Abundance of deuterium in high red-shift Lyman alpha absorption systems
}
\label{t-h2}
\newcommand{\m}{\hphantom{$-$}}
\newcommand{\cc}[1]{\multicolumn{1}{c}{#1}}
\renewcommand{\arraystretch}{1.4} 
\begin{tabular}{@{}lllllllll}
\hline
 $z$ & \cc{$3.514$}\cite{7} & \cc{$2.0762$}\cite{5} & \cc{$3.025$}\cite{4} 
& \cc{$2.536$}\cite{3} & 
  \cc{$3.025$}\cite{8} & \cc{$3.572$}\cite{2} &
\cc{$2.504$}\cite{1} &  \cc{$0.701$}\cite{6}
\\
\hline
 $10^5(D/H)$ &$\sim$ 1.5 & $1.65\pm 0.35$ &$ 2.24\pm 0.67$ 
& $2.54\pm 0.23$  & 
$ 3.2\pm 0.4$ & $3.25\pm 0.3$ & $4.0\pm 0.65$ 
& $20\pm 5$ 
\\
\hline
\end{tabular}
\end{table*}

The entries in Table~\ref{t-h2} are situated in order of increasing the
$(D/H)$-values. Entries in columns 3 and 5 correspond to the same system
but the result was changed to a larger value after its reanalysis in 
ref.~\cite{8} where a complex velocity structure suggested by S II and 
$H_2$ absorption lines was taken into account. 

The data seem to be grouped around three central points.
The majority of the data are accumulated around $10^5\,(D/H) = 3$. 
There are two measurements indicating low deuterium about 1.5 and one
showing a huge value, $20\pm 5$. Taken by the face value, there is an
indication that there are systems with high, normal,
and low deuterium abundances. If this is true, it would be a confirmation
of the model of ref.~\cite{dolgov99} where spatial variation of primordial
abundances are predicted. According to this work, two thirds of the sky
should have normal abundance of deuterium, 1/6 should have twice smaller
ratio of $D/H$, and another 1/6 should have about 5 times larger than
normal abundance of deuterium. On the other hand, as is argued in 
ref.~\cite{kirkman01}, the line observed in the $z=0.701$
absorption system~\cite{6} which indicates a high deuterium abundance
most probably is not deuterium. This statement is based on a new 
measurement of the velocity dispersion of the neutral hydrogen.

\subsection{Observations of helium-4 \label{ss-he4}}

Helium-4 is very strongly binded nuclei so it cannot be destroyed
by stellar processes but only produced in stars together with other heavier
elements, ``metals'', e.g. oxygen (O) and nitrogen (N) - all elements 
heavier than 
$^4 He$ are called ``metals'' in astronomy. Thus the observed mass
fraction of $^4He$, should be larger than the primordial one, $Y_p$.
$^4 He$ is observed in hot ionized gas regions (H II) in emission optical
recombination lines. In contrast to deuterium $^4 He$ has been observed 
only at relatively small distances, maximum, $z= 0.045$~\cite{izotov97}.
This regions are contaminated by stellar processes and to infer primordial 
abundance from the data one should extrapolate to zero metallicity, i.e.
to zero abundances of O or N. A detailed discussion of these issues can 
be found e.g. in the book~\cite{pagel97}. The extrapolation to zero 
metallicity is realized by the linear function with the slope
$\kappa = dY/dZ$ where $Y$ is the observed mass fraction of $^4He$ and
$Z$ is that of metals. Analysis by different groups give significantly
different values $\kappa = 6.7\pm 2.3$~\cite{pagel92},
$6.9\pm 1.5$~\cite{olive97}, and $\kappa=2.4\pm 1.0$~\cite{thuan00}. 
Though $\kappa$ is not accurately determined (or it may have different
values for different systems), the systems where $^4He$ is studied have
a low metallicity and the net effect on primordial abundance of $^4 He$
is not very strong.
 
The values of primordial mass fraction
of $^4 He$, $Y_p$, measured by several different groups are presented
in table~\ref{t-he4}. The discrepancy between the results can be
possibly prescribed to different treatment of correction factors:
1) ionization correction factor (icf) which determines how much neutral 
hydrogen (invisible) is in the object under scrutiny, 2) temperature
correction factor (tcf) which describes non-uniform temperature 
distribution, and 3) density of electrons. A recent analysis of the
correction has been performed in ref.~\cite{sauer01}, where it was
shown in particular that combined icf and tcf would diminish the result by 
0.002-0.004. It is worth noticing in this connection that the reanalysis
of the data of ref.~\cite{thuan00} in the paper~\cite{gruenwald01}
where icf has been taken into account, has led to a noticeably smaller
result (see table~\ref{t-he4}).

\begin{table*}[htb]
\caption{
Primordial mass fraction of $^4 He$ according to the results of 
different groups.
}
\label{t-he4}
\newcommand{\m}{\hphantom{$-$}}
\newcommand{\cc}[1]{\multicolumn{1}{c}{#1}}
\renewcommand{\arraystretch}{1.4} 
\begin{tabular}{@{}lllllll}
\hline
 Ref. &  \cite{pagel92} & \cite{olive97} &  \cite{peimbert01} 
& \cite{thuan00} & \cite{gruenwald01} 
\\
\hline
 $Y_p$ & $ 0.228\pm 0.005 $  & $ 0.234\pm 0.002$ &$ 0.2348\pm 0.0025$ 
& $0.2443\pm 0.0015$  & 
$ 0.238\pm 0.003$ 
\\
\hline
\end{tabular}
\end{table*}

Even the largest mass fraction of $^4 He$ presented in table~\ref{t-he4} 
corresponds to a rather low baryon number density, 
$\eta_{10} \approx 4.2$ which is noticeably smaller than the value
determined from CMBR~(\ref{omegab}). Possibly it means that systematic
uncertainties and correction factors are underestimated and more work is 
necessary to obtain more precise results. This problem is discussed 
in the recent paper~\cite{olive00s}. It seems that at the present time
the accuracy in determination of primordial mass fraction of $^4 He$
is not sufficiently good.

\section{BIG BANG NUCLEOSYNTHESIS AND NON-STANDARD PHYSICS \label{s-nonst}}

Many possible modifications of the standard cosmological scenario and/or
Minimal Standard Model (MSM) in elementary particle physics can be strongly
constrained or even excluded by BBN. As we have already mentioned, BBN 
permits to restrict the number of neutrino families. The most recent 
analysis~\cite{cyburt01}, based on $Y_p = 0.238\pm 0.002 \pm 0.005$, 
results in the following 95\% CL upper limits:
\be
\Delta N < \left\{ \begin{array}{ll}
0.6 & \mbox{{\rm for}\,\,\,$\eta_{10} =5.8$ },  \\
0.9 & \mbox{{\rm for}\,\,\,$\eta_{10} =2.4 $},
 \end{array}\right.
\label{delta-n}
\ee
though in an earlier paper~\cite{tytler00} a much more restrictive bound, 
$\Delta N < 0.2$, was advocated. At the present stage is safe to assume 
that $\Delta N <1$. Hopefully in the near future one will be able to derive
a stronger limit.

Other additional parameters that can influence light element abundances 
are chemical potentials of different neutrino species, 
$\mu_a$, where $a=e,\mu,\tau$.
A possible role of neutrino degeneracy in big bang nucleosynthesis was
noted already in the pioneering work by Wagoner, Fowler, and
Hoyle~\cite{wagoner67} and after that it was analyzed in a number of
papers. A combined analysis of the effect of simultaneous variation of 
all three chemical potentials on BBN was recently performed in the
papers~\cite{orito00,esposito00b}. Since the effect of positive $\mu_e$,
that diminishes $n/p$-ratio, can be compensated by a non-zero 
$\mu_\mu$ or
$\mu_\tau$, the limits obtained without any extra information are rather
loose, assuming that the mentioned above conspiracy between 
$\mu_e$ and $\mu_{\mu,\tau}$ exists. However, additional data on the 
angular spectrum of CMBR permits to obtain~\cite{hansen01} stronger bounds
even if the conspiracy is allowed:
\be
-0.01 <\xi_{\nu_e} < 0.2,\,\,\,\, |\xi_{\nu_\mu,\nu_\tau}| < 2.6
\label{limitsxi}
\ee
These results are derived
under assumptions that primordial fraction of deuterium is
$D/H = (3.0 \pm 0.4)\cdot 10^{-5}$. Here $\xi_a = \mu_a/T$ are 
dimensionless chemical potentials.

These bounds are valid in absence of neutrino oscillations. On the other
hand, solar and atmospheric neutrino anomalies present a strong evidence
in favor of mixing between $\nue$, $\num$, and $\nut$ (see e.g. talks
by Y. Totsuka and  A. McDonald at this Conference). Oscillations of these 
neutrinos do not conserve individual leptonic charges and a primordial
lepton asymmetries in electronic, muonic, and tauonic charges would be 
redistributed by the oscillations. In particular, a large $\mu_\mu$ or 
$\mu_\tau$ would create a noticeable $\mu_e$ of the same sign. An analysis
of active-active neutrino oscillations have not yet been accurately
performed. Preliminary results~\cite{dolgov01} are $|\xi_a| < 0.1$ for
any flavor $a=e,\mu,\tau$. Let us note that in the case of vanishing
chemical potentials oscillations between active neutrinos would not have
a noticeable impact on BBN, because the neutrinos would practically remain 
in thermal equilibrium independently of oscillations and no deviation from 
the standard scenario would emerge. There could be a small effect due to 
spectral distortion of neutrinos by late 
$e^+e^-$-annihilation~\cite{dolgov92}. Oscillations could change the 
distorted spectrum and might be in principle observable in BBN.

It is still not excluded that sterile companions of active neutrinos
exist. Moreover, recent measurements~\cite{aguilar01}
seem to confirm creation of $\bar\nue$
in $\bar\num$ beam. If this is the case, an interpretation of all neutrino 
data in terms of oscillations demands an existence of one (or several)
sterile neutrino(s), $\nu_s$, mixed with active ones. 

In contrast to equilibrium
active-active case, mixing of active and sterile neutrinos should have a 
noticeable impact on BBN. There are three possible effects due to 
oscillations: 1) production of new relativistic states, $\nu_s$, and an
increase of $N_\nu$; 2) distortion of neutrino spectrum, especially 
important is $\nue$ spectrum; 3) resonance generation of a large lepton
asymmetry from a very small initial leptonic or baryonic 
ones~\cite{foot96} (only the pioneering and most recent papers are quoted; 
references to many other ones can be found therein). 

The effect of creation of additional relativistic species 
by oscillations is easily estimated 
and from the condition that BBN allows $\Delta N$ extra
neutrino families one can obtain the bound on vacuum mixing angle and
the mass difference:
\be
(\delta m^2/{\rm eV^2})\,\sin^4 2\theta <\left\{ \begin{array}{ll}
3.16\cdot 10^{-5} (\Delta N)^2 & 
  \\
1.74\cdot 10^{-5} (\Delta N)^2 & 
 \end{array}\right.
\label{dmsin}
\ee
for $\nue\,\nu_s$ and $\nu_{\mu,\tau}\,\nu_s$ mixings respectively.
This bounds are meaningful only if $\Delta N <1$.

Discussion of the effects of neutrino spectrum distortion on BBN can be 
found in ref.~\cite{kirilova01}. The impact of asymmetry generation on
BBN is discussed in ref.~\cite{foot96} (second paper) and in
refs.~\cite{foot99}. Resonance case is quite complicated and the effect
may have either sign. No simple bounds can be presented here.

BBN permits put stringent bounds on possible time-variation of fundamental 
constants. If we assume e.g. that the fine structure constant $\alpha$
is different at BBN from its present-day value, $\alpha = 1/137$, we should
expect that the neutron-proton mass difference should also change with time.
The $(n-p)$-mass difference is given by
$\delta m \equiv m_n-m_p = m_d -m_u +\alpha m_{em}$,
where $m_d$ and $m_u$ are the masses of $u$ and $d$ quarks and the last term
describes electromagnetic loop contribution into $\delta m$. As we have seen
above the $n/p$-ratio is equal to $n/p = \exp (-\delta m/T_f)$, where $T_f$ 
is the freezing temperature of reactions~(\ref{np-react}). This temperature
is determined by the competition between the Hubble expansion and
weak interaction rates. The latter is proportional to the magnitude of
the electroweak coupling constant, which is essentially $\alpha$. Hence
$T_f \sim \alpha^{-2/3}$. Demanding that successful results of BBN are not
destroyed one can obtain that 
$(\Delta \alpha /\alpha)_{BBN} < $ (a few)$\times 10^{-2}$.

\section{Conclusion}

We see that gross features of BBN well agree with observations but 
the latter are not yet sufficiently accurate to make it really a precise 
science. Moreover there is a trend to discrepancy between the observations 
of deuterium which indicate a higher value of $\eta_{10}$ than the
observations of $^4 He$. Hopefully it will be clear in a few coming years
if this is a real problem or an artifact of systematic and evolutionary
effects. We have not discussed above primordial $^7 Li$ because the
accuracy of its measurements are rather low now but potentially this
element could be very important for verification of the standard model.
A recent discussion of $^7 Li$ can be found in ref.~\cite{olive00}.
Still even with the existing level of accuracy BBN permit to put powerful 
constraints on deviations from the standard model. The number of extra 
neutrino species allowed by the contemporary observations 
is about unity and with this bound very little can
said about mixing parameters between active and sterile neutrinos.
However, if $\Delta N$ could be 
reduced, say, to 0.3 the limit would be meaningful. The 
restriction on the time $\alpha$ is quite strong and may exclude some of
the models predicting such variation.

\end{document}